\documentclass[runningheads,envcountsame,citeauthoryear]{llncs}
\usepackage[english]{babel}
\usepackage[latin1]{inputenc}
\usepackage[T1]{fontenc}

\usepackage{cc4llncs}

\usepackage{mathell}
\usepackage{set}
\usepackage{weakhyph}
\usepackage{cyrillic}
\usepackage{graphicx}

\newcommand{\person}[1]{#1}
\newcommand{\myomega}{\omega_{2}}

\newcommand{\bigO}{\operatorname{\mathcal{O}}}
\newcommand{\softO}{\operatorname{\mathcal{O}^{\sim}}}
\newcommand{\smoothO}{\operatorname{\mathcal{O}^{\approx}}}
\newcommand{\smallo}{\operatorname{o}}
\newcommand{\loglog}[1][]{%
  \operatorname{\textnormal{log}_{#1}\,\textnormal{log}_{#1}}\nolimits}
\newcommand{\maxdeg}{\operatorname{max\,deg}}
\newcommand{\floor}[1]{\left\lfloor#1\right\rfloor}
\newcommand{\ceil}[1]{\left\lceil#1\right\rceil}
\newcommand{\rem}{\mathbin{\operatorname{rem}}}
\newcommand{\n}{m}  

\newcommand{\bbbR}{\mathbb{R}}
\newcommand{\bbbN}{\mathbb{N}}

\newcommand{\bbbC}{\mathbb{C}}

\newcommand{\bbbK}{\mathbb{K}}
\newcommand{\bbbL}{\mathbb{L}}

\newcommand{\cf}{see}
\newcommand{\eg}{for example}
\newcommand{\ie}{that is}
\newcommand{\innset}[2]{0\leq #1 < #2}

\let\epsilon\varepsilon
\let\phi\varphi

\begin{document}

\title{Fast Multipoint Evaluation\\ of Bivariate Polynomials}

\author{Michael N\"{u}sken \and Martin Ziegler\thanks{Supported by the
    DFG Research Training Group \textsf{GK-693} of the Paderborn
    Institute for Scientific Computation (\textsf{PaSCo})}}
\institute{University of Paderborn, 33095 Paderborn, GERMANY \\
  \email{\{nuesken,ziegler\}@upb.de}}

\date{June 22, 2004}

\maketitle

\begin{abstract}
  We generalize univariate multipoint evaluation of polynomials of
  degree $n$ at sublinear amortized cost per point.  More precisely,
  it is shown how to evaluate a bivariate polynomial $p$ of maximum
  degree less than $n$, specified by its $n^{2}$ coefficients,
  simultaneously at $n^{2}$ given points using a total of
  $\bigO(n^{2.667})$ arithmetic operations. In terms of the input size
  $N$ being quadratic in $n$, this amounts to an amortized cost of
  $\bigO(N^{0.334})$ per point.
\end{abstract}

\section{Introduction}

By \person{Horner}'s Rule, any polynomial $p$ of degree less than $n$
can be evaluated at a given argument $x$ in $\bigO(n)$ arithmetic
operations which is optimal for a generic polynomial as proved by
\cite{pan66}, \cf{} \eg{} Theorem~6.5 in \citet*{burcla97}.

In order to evaluate $p$ at \emph{several} points, we might
sequentially compute $p(x_{k})$ for $0 \leq k < n$.  However,
regarding that both the input consisting of $n$ coefficients of $p$
and $n$ points $x_{k}$ and the output consisting of the $n$ values
$p(x_{k})$ have only linear size, information theory provides no
justification for this quadratic total running time.  In fact, a more
sophisticated algorithm permits to compute all $p(x_{k})$
simultaneously using only $\bigO(n \cdot \log^{2} n \cdot \loglog n)$
operations.  Based on the \emph{Fast \person{Fourier} Transform}, the
mentioned algorithms and others realize what is known as \emph{Fast
  Polynomial Arithmetic}.
For ease of notation, we use the `soft-Oh' notation, namely
$\softO(f(n)) := \bigO \left( f(n) (\log f(n))^{\bigO(1)} \right)$.
This variant of the usual asymptotic `big-Oh' notation ignores
poly-logarithmic factors like $\log^{2} n \cdot \loglog n$.

\begin{fact}
  \label{Unibasics}
  Let $R$ be a commutative ring with one.
  \begin{enumerate}
  \item\label{Unibasics:a} \emph{Multiplication of univariate
      polynomials}: Suppose we are given polynomials $p, q \in R[X]$
    of degree less than $n$, specified by their coefficients. Then we
    can compute the coefficients of the product polynomial $p\cdot q
    \in R[X]$ using $\softO(n)$ arithmetic operations in $R$.
  \item\label{Unibasics:b} \emph{Multipoint evaluation of a univariate
      polynomial}: Suppose we are given a polynomial $p \in R[X]$ of
    degree less than $n$, again specified by its coefficients, and
    points $x_{0},\ldots,x_{n-1}\in R$.  Then we can compute the
    values $p(x_{0}),\ldots,p(x_{n-1})\in R$ using $\softO(n)$
    arithmetic operations in $R$.
  \item\label{Unibasics:c} \emph{Univariate interpolation}:
    Conversely, suppose we are given points $(x_{k},y_{k}) \in R^{2}$
    for $0 \leq k < n$ such that $x_{k}-x_{\ell}$ is invertible in $R$
    for all $k\ne\ell$.  Then we can compute the coefficients of
    a 
    polynomial $p\in R[X]$ of degree less than $n$ such that
    $p(x_{k})=y_{j}$, $0 \leq k < n$, \ie{}, determine the
    interpolation polynomial to data $(x_{k},y_{k})$ using $\softO(n)$
    arithmetic operations in $R$.
  \end{enumerate}
\end{fact}
\begin{proof}
  These results can be found \eg{} in \cite{gatger03} including small
  constants:
  \begin{itemize}
  \item[\ref{Unibasics:a}] can be done using at most $63.427 \cdot
    n \cdot \log_{2} n \cdot \loglog[2] n + \bigO(n \log n)$
    arithmetic operations in $R$ by Theorem~8.23.  The essential
    ingredient is the Fast Fourier Transform.  If $R = \bbbC$ then
    even $\frac{9}{2} n \log_{2} n + \bigO(n)$ arithmetic operations
    suffice. This goes back to \cite{schstr71} and \cite{sch77}.
  \end{itemize}
  In the following $\textsf{M}(n)$ denotes the cost of one
  multiplication of univariate polynomials over $R$ of degree less
  then $n$.
  \begin{itemize}
  \item[\ref{Unibasics:b}] can be done using at most $\frac{11}{2}
    \textsf{M}(n) \log_{2} n + \bigO(n \log n )$ operations in $R$
    according to Corollary~10.8.  Here, Divide \&{} Conquer provides
    the final building block.  This goes back to \cite{fid72}.
  \item[\ref{Unibasics:c}] can be done using at most $\frac{13}{2}
    \textsf{M}(n) \log_{2} n + \bigO(n \log n )$ operations in $R$
    according to Corollary~10.12.  This, too, is completed by Divide
    \&{} Conquer.  The result goes back to \cite{hor72}.
  \end{itemize}
  You also find an excellent account of all these in \cite{bormun75}.
  \qed%
\end{proof}
Fast polynomial arithmetic and in particular multipoint evaluation has
found many applications in algorithmic number theory \citep[\cf{}
\eg{}][]{odlsch88}, computer aided geometric design \citep[\cf{}
\eg{}][]{lodgol97}, and computational physics \citep[\cf{}
\eg{}][]{zie03a}.

Observe that the above claims apply to the \emph{univariate} case.
What about multivariate analogues?  Let us for a start consider the
bivariate case: A bivariate polynomial $p \in R[X,Y]$ of \emph{maximum
  degree} $\maxdeg p := \max\Set{\deg_{X}p,\deg_{Y}p}$ less than $n$
has up to $n^{2}$ coefficients, one for each monomial $X^{i} Y^{j}$
with $\innset{i, j}{n}$.
Now corresponding to \ref{Unibasics}, the following questions emerge:
\begin{question}
  \begin{enumerate}
    \let\origitem\item
    \renewcommand{\item}[1][]{\origitem{}\emph{#1}:}
  \item [Multiplication of bivariate polynomials]
    \label{Bibasics:a}
    Can two given bivariate polynomials of maximum degree less than
    $n$ be multiplied within time $\softO(n^{2})$?
  \item [Multipoint evaluation of a bivariate polynomial]
    \label{Bibasics:b}
    Can a given bivariate polynomial of maximum degree less than $n$
    be evaluated simultaneously at $n^{2}$ arguments in time
    $\softO(n^{2})$?
  \item [Bivariate interpolation]
    \label{Bibasics:c}
    Given $n^{2}$ points $(x_{k},y_{k},z_{k}) \in R^{3}$, is there a
    polynomial $p \in R[X,Y]$ of maximum degree less than $n$ such
    that $p(x_{k},y_{k})=z_{k}$ for all $0 \leq k < n^{2}$?  And, if
    yes, can we compute it in time $\softO(n^{2})$?
  \end{enumerate}
\end{question}
Such issues also arise for instance in connection with fast arithmetic
for polynomials over the skew-field of hypercomplex numbers
\citep[Section~3.1]{zie03b}.

A positive answer to \whole\ref{Bibasics:a} is achieved by embedding
$p$ and $q$ into univariate polynomials of degree $\bigO(n^{2})$ using
the \emph{\person{Kronecker} substitution} $Y = X^{2n-1}$, applying
\ref{Unibasics:a} to them, and then re-substituting the result to a
bivariate polynomial; \cf{} \eg{} Corollary~8.28 in \cite{gatger03} or
Section~1.8 in \cite{binpan94}.

Note that the first part of \ref{Bibasics:c} has negative answer for
instance whenever the points $(x_{k},y_{k})$ are co-linear or, more
generally, lie on a curve of small degree: Here, a bivariate
polynomial of maximum degree less than $n$ does not even exist in
general.

Addressing \ref{Bibasics:b}, observe that \person{Kronecker}
substitution is not compatible with evaluation and thus of no direct
use for reducing to the univariate case.  The methods that yield
\ref{Unibasics:b} are not applicable either as they rely on fast
polynomial division with remainder which looses many of its nice
mathematical and computational properties when passing from the
univariate to the bivariate case.

Nevertheless, \ref{Bibasics:b}~does admit a rather immediate positive
answer \emph{provided} the arguments $(x_{k},y_{k})$, $0 \leq k <
n^{2}$ form a \person{Cartesian} $n\times n$-grid (also called tensor
product grid).  Indeed, consider $p(X,Y) = \sum_{0\leq j < n} q_{j}(X)
Y^{j}$ as a polynomial in $Y$ with coefficients $q_{j}$ being
univariate polynomials in $X$.  Then multi-evaluate $q_{j}$ at the $n$
distinct values $x_{k}$: as $q_{j}$ has degree less than $n$, this
takes time $\softO(n)$ for each $j$, adding to a total of
$\softO(n^{2})$.  Finally take the $n$ different univariate
polynomials $p(x_{k},Y)$ in $Y$ of degree less than $n$ and
multi-evaluate each at the $n$ distinct values $y_{\ell}$: this takes
another $\softO(n^{2})$.

The presumption on the arguments to form a \person{Cartesian} grid
allows for a slight relaxation in that this grid may be rotated and
sheared:
\begin{figure}[h]
  \centering \includegraphics[width=\textwidth]{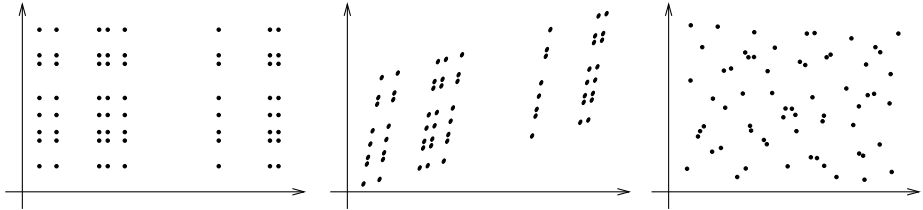}
  \caption{%
    Cartesian $8\times8$-grid, \qquad same rotated and sheared; \qquad
    64 generic points.}
\end{figure}
Such kind of affine distortion is easy to detect, reverted to the
arguments, and then instead applied to the polynomial~$p$ by
transforming its coefficients within time $\softO(n^{2})$, see
\ref{Subst} below.  The obtained polynomial $\hat{p}$ can then be
evaluated on the now strictly \person{Cartesian} grid as described
above.  However, $n\times n$ grids, even rotated and sheared ones,
form only a zero-set within the $2n^{2}$-dimensional space of all
possible configurations of $n^{2}$ points.  Thus this is a severe
restriction.

\section{Goal and Idea}
\label{secIdea}

The big open question and goal of the present work is concerned with
fast multipoint evaluation of a multivariate polynomial.  As a first
step in this direction we consider the bivariate case.

The na{\"{\i}}ve approach to this problem, namely of sequentially
calculating all $p(x_{k},y_{k})$, takes quadratic time each, thus
inferring total cost of order $n^{4}$.  A first improvement to
$\softO(n^{3})$ is based on the simple observation that any $n$~points
in the plane can easily be extended to an $n\times n$ grid on which,
by the above considerations, multipoint evaluation of $p$ is feasible
in time $\softO(n^{2})$.  So we may partition the $n^{2}$ arguments
into $n$ blocks of $n$ points and multi-evaluate $p$ sequentially on
each of them to obtain the following 
\begin{theorem}
  \label{Gridextension}
  Let $R$ be a commutative ring with one.
  A bivariate polynomial $p\in R[X,Y]$ of $\deg_X(p)<n$ and
  $\deg_Y(p)<n$, given by its coefficients, can be evaluated
  simultaneously at $n^2$ given arguments $(x_k,y_k)$ using at most
  $\bigO(n^3\cdot\log^2n\cdot\loglog n)$ arithmetic operations in $R$.
\end{theorem}

We reduce this softly cubic upper complexity bound to
$\bigO(n^{2.667})$. More precisely, by combining fast univariate
polynomial arithmetic with fast matrix multiplication we will prove:
\begin{result} 
  \label{Main}
  Let $\bbbK$ denote an arbitrary field.
  A bivariate polynomial $p\in\bbbK[X,Y]$ of $\deg_{X}(p)<n$ and
  $\deg_{Y}(p)<m$, specified by its coefficients, can be evaluated
  simultaneously at $N$ given arguments $(x_k,y_k)\in\bbbK^2$ with
  pairwise different first coordinates using $\bigO\left((N+n m)
    m^{\myomega/2-1+\epsilon}\right)$ arithmetic operations in $\bbbK$
  for any fixed $\epsilon>0$.
\end{result}
Here, $\myomega$ denotes the exponent of the multiplication of
$n\times n$- by rectangular $n\times n^2$-matrices, \cf{}
Section~\ref{secMatMul}. In fact this problem is well-known to admit a
much faster solution than na{\"{\i}}ve $\bigO(n^4)$, the current world
record $\myomega<3.334$ being due to \cite{huapan98}.  By choosing
$m=n$ and $N=n^2$, this yields the running time claimed in the
abstract.

The general idea underlying \ref{Main}, illustrated for the case of
$n=m$, is to reduce the bivariate to the univariate case by
substituting $Y$ in $p(X,Y)$ with the interpolation polynomial $g(X)$
of degree less than $n^{2}$ to data $(x_{k},y_{k})$.  It then suffices
to multi-evaluate the univariate result $p\big(X,g(X)\big)$ at the
$n^{2}$ arguments $x_{k}$.  Obviously, this can only work if such an
interpolation polynomial $g$ is available, \ie{} any two evaluation
points $(x_{k},y_{k}) \ne (x_{k'},y_{k'})$ differ in their first
coordinates, $x_{k} \ne x_{k'}$.  However, this condition can be
asserted easily later on, see Section~\ref{secRestriction},
so for now assume it is fulfilled.

This na{\"{\i}}ve substitution leads to a polynomial of degree up to
$\bigO(n^{3})$.  On the other hand, it obviously suffices to obtain
$p\big(X,g(X)\big)$ modulo the polynomial $f(X) := \prod_{0 \leq k <
  n^{2}} (X-x_{k})$ which has degree less than $n^{2}$.  The key to
efficient bivariate multipoint evaluation is thus an efficient
algorithm for this \emph{modular bi-to-univariate composition}
problem, presented in \ref{Comp}.

As we make heavy use of fast matrix multiplication,
Section~\ref{secMatMul} recalls some basic facts, observations, and
the state of the art in that field of research.
Section~\ref{secResults} formally states the main result of the
present work together with two tools (affine substitution and modular
composition) which might be interesting on their own, their proofs
being postponed to Section~\ref{secProofs}.
Section~\ref{secRestriction} describes three ways to deal with
arguments that do have coinciding first coordinates.
Section~\ref{secConclusion} gives some final remarks.

\section{Basics on Fast Matrix Multiplication}
\label{secMatMul}

Recall that, for a field $\bbbK$, $\omega = \omega(\bbbK)\geq 2$
denotes the \emph{exponent of matrix multiplication}, \ie{}, the least
real such that $\n \times \n$ matrix multiplication is feasible in
asymptotic time $\bigO(\n^{\omega+\epsilon})$ for any $\epsilon>0$;
\cf{} \eg{} Chapter~15 in \cite{burcla97}.  The current world-record
due to \cite{copwin90} achieves $\omega<2.376$ independent of the
ground field $\bbbK$.  The Notes~12.1 in \cite{gatger03} contain a
short historical account.

Clearly, a rectangular matrix multiplication of, say, $\n \times
\n$-matrices by $\n \times \n^{t}$-matrices can always be done
partitioning into $\n \times \n$ square matrices.  Yet, in some cases
there are better known algorithms than this.
We use the notation introduced by \cite{huapan98}: $\omega(r,s,t)$
denotes the exponent of the multiplication of $\ceil{\n^{r}} \times
\ceil{\n^{s}}$- by $\ceil{\n^{s}} \times \ceil{\n^{t}}$-matrices,
\ie{}
$$
\omega(r,s,t) = \inf \Set{\tau \in \bbbR;
  \parbox{15em}{Multiplication of $\ceil{\n^{r}} \times
    \ceil{\n^{s}}$- by $\ceil{\n^{s}} \times \ceil{\n^{t}}$-matrices
    can be done with $\bigO(\n^{\tau})$ arithmetic operations}}.
$$
Clearly, $\omega = \omega(1,1,1)$.  We always have
\begin{equation} \label{eqOmega}
  \max\Set{r+s,r+t,s+t} \quad\leq\quad \omega(r,s,t) \quad\leq\quad r+s+t.
\end{equation}
Note that $\omega(r,s,t)$ is in fact invariant under permutation of
its arguments.

We collect some known bounds on fast matrix multiplication algorithms.
\begin{fact}
  \begin{enumerate}
  \item\label{MM:a} $\omega = \omega(1,1,1) \leq \log_{2}(7) <
    2.8073549221$ \citep{str69}.
  \item\label{MM:b} $\omega = \omega(1,1,1) <
    2.3754769128$ \citep{copwin90}.
  \item\label{MM:c} $\myomega := \omega(1,1,2) <
    3.3339532438$ \citep{huapan98}.
  \end{enumerate}
\end{fact}
Partitioning into square matrices only yields $\myomega \leq \omega+1
< 3.3754769128$.  Bounds for further rectangular matrix
multiplications can be also be found in \cite{huapan98}.  It is
conjectured that $\omega = 2$.  Then by partitioning into square
blocks also $\omega(r,s,t)$ touches its lower bound
in~(\ref{eqOmega}), \ie{} $\omega(r,s,t) = \max\Set{r+s,r+t,s+t}$.  In
particular, $\myomega=3$ then.

We point out that the definition of $\omega$ and $\omega(r,s,t)$
refers to arbitrary algebraic computations which furthermore may be
non-uniform, that is, use for each matrix size $\n$ a different
algorithm.  However, closer inspection of Section~15.1 in
\cite{burcla97} reveals the following
\begin{observation}
  \label{MM:d}
  Rectangular matrix multiplication of $\ceil{\n^{r}} \times
  \ceil{\n^{s}}$- by $\ceil{\n^{s}} \times \ceil{\n^{t}}$-matrices
  over $\bbbK$ can be done with $\bigO(\n^{\omega(r,s,t)+\epsilon})$
  arithmetic operations in $\bbbK$ by a uniform, bilinear algorithm
  for any fixed $\epsilon$.
\end{observation}
A bilinear computation is a very special kind of algorithm where apart
from additions and scalar multiplications only bilinear
multiplications occur; see \eg{} Definition~14.7 in \cite{burcla97}
for more details.  In particular, no divisions are allowed.

\section{Main results}
\label{secResults}

Our major contribution concerns bivariate multi-evaluation at
arguments $(x_k,y_k)$ under the condition that their first coordinates
$x_k$ are pairwise distinct. This amounts to a weakened \emph{general
  position} presumption as is common for instance in Computational
Geometry.

For notational convenience, we define `$\smoothO$' (smooth-Oh) which,
in addition to polylogarithmic factors in $n$, also ignores factors
$n^\epsilon$ as long as $\epsilon>0$ can be chosen arbitrarily small.
Formally, $\smoothO(f(n)) := \bigcap_{\epsilon>0} \bigO(
f(n)^{1+\epsilon})$.  Note that $\softO(f(n)) \subset \smoothO(f(n))$.

\begin{theorem}
  \label{Main2}
  Let $\bbbK$ denote a field.
  Suppose $n, m \in \bbbN$.  Given the $n m$ coefficients of a
  bivariate polynomial $p$ with $\deg_{X}(p) < n$ and $\deg_{Y}(p)<m$
  and given $n m$ points $(x_{k},y_{k})\in\bbbK^2$, ${\innset{k}{n
      m}}$ such that the first coordinates $x_{k}$ are pairwise
  different, we can calculate the $n$ values $p(x_{k},y_{k})$ using
  $\smoothO\left( n m^{\myomega/2} \right)$ arithmetic operations over
  $\bbbK$.
  The algorithm is uniform.
\end{theorem}

Observe that this yields the first part of \ref{Main} by performing
$\left\lceil N/(n m) \right\rceil$ separate multipoint evaluations at
$n m$ points each.  Let us also remark that any further progress in
matrix multiplication immediately carries over to our problem.  As it
is conjectured that $\omega=2$ holds, this would lead to bivariate
multipoint evaluation within time $\smoothO(n m^{1.5})$.

\medskip

Our proof of \ref{Main2} is based on the following generalization of
\citeauthor{brekun78}s efficient \emph{univariate} modular
composition, \cf{} \eg{} Section~12.2 in \cite{gatger03},
to a certain `\emph{bi-to-univariate}' variant:
\begin{theorem}
  \label{Comp}
  Fix a field $\bbbK$.
  Given $n,m\in\bbbN$, a bivariate polynomial $p\in\bbbK[X,Y]$ with
  $\deg_{X}(p) < n$ and $\deg_{Y}(p)<m$ and univariate polynomials $g,
  f \in \bbbK[X]$ of degree less than $n m$, specified by their
  coefficients.  Then $p\big(X,g(X)\big)\rem f(X)$ can be computed
  with $\smoothO(n m^{\myomega/2})$ arithmetic operations in $\bbbK$.
\end{theorem}
We remark that \emph{true} bivariate modular computation requires
\person{Gr\"{o}bner} basis methods which for complexity reasons are
beyond our interest here.

\pagebreak
\section{Proofs}
\label{secProofs}

Now we come to the proofs.


\begin{lemma}
  \label{Matrix}
  Let $\bbbK$ denote a field and fix $t>0$.
  \begin{enumerate}
  \item
    \label{Matrix:a}
    Let both $A$ be an $m \times m$-matrix and $B$ an $m \times
    m^{t}$-matrix whose entries consist of polynomials $a_{ij}(X),
    b_{ij}(X) \in \bbbK[X]$ of degree less than $n$.  Given $m$ and
    the $n\cdot\left( m^{2} + m^{t} \right)$ coefficients, we can
    compute the coefficients of the polynomial entries $c_{ij}(X)$ of
    $C:=A\cdot B$ within $\smoothO(n m^{\omega(1,1,t)})$ arithmetic
    operations.
  \item
    \label{Matrix:d}
    If $A$ denotes an $m\times m$ square matrix with polynomial
    entries of degree less than $n$ and $b$ denotes an $m$-component
    vector of polynomials of degree less than $n m^{t}$, then $(A,b)
    \mapsto A\cdot b$ is computable within $\smoothO(n
    m^{\omega(1,1,t)})$.
  \item
    \label{Comp:a}
    Let $p_{0}, \ldots, p_{m-1} \in \bbbK[X,Y]$ denote bivariate
    polynomials with $\deg_{X}(p_{i}) < n$ and $\deg_{Y}(p_{i}) < m$,
    given their $n m^{2}$ coefficients, and let furthermore univariate
    polynomials $g, f \in \bbbK[X]$ of degree less than $n m^{t}$ be
    given by their coefficients.  Then the coefficients of the $m$
    univariate polynomials
    $$
    p_{i}\big(X,g(X)\big) \rem f(X)
    $$
    can be computed with $\smoothO(n m^{\omega(1,1,t)})$ arithmetic
    operations.
  \end{enumerate}
  In particular, for $t=1$ we have cost $\smoothO(n m^{\omega})
  \subset \softO(n m^{2.376})$ and for $t=2$ we have cost $\smoothO(n
  m^{\myomega}) \subset \softO(n m^{3.334})$.
\end{lemma}

\begin{proof}
  \begin{enumerate}
  \item[\ref{Matrix:a}] %
    By scalar extension to $R = \bbbK[X]$ we obtain an algorithm with
    cost $\smoothO(m^{\omega(1,1,t)})$ arithmetic operations in $R$
    using \ref{MM:d}.  For the algorithm scalar extension simply means
    that we perform any multiplication in $R$ instead of $\bbbK$,
    multiplications with constants become scalar multiplications.  And
    the cost for one operation in $R$ is $\softO(n)$ as only
    polynomials of degree $n$ have to be multiplied.
  \item[\ref{Matrix:d}] For each $j$, $0 \leq j < m$, decompose the
    polynomial $b_{j}$ of degree less than $n m^{t}$ into $m^{t}$
    polynomials of degree less than $n$, \ie{}, write $b_{j}(X) =
    \sum_{0\leq k<m^{t}} b_{jk}(X) \cdot X^{kn}$.  The desired
    polynomial vector is then given by
    \begin{equation}
      \tag{$*$}
      \label{eqSum}
      \begin{aligned}
        \big(A\cdot b\big)_{i}(X)
        =&%
        \sum_{1\leq j\leq m} a_{ij}(X)\cdot\left( \smash{\sum_{0 \leq
              k<m^{t}}}\vphantom{\sum} b_{jk}(X)\cdot X^{kn}\right)
        \\
        =& \sum_{0 \leq k < m^{t}} \big(A\cdot B\big)_{ik}(X) \cdot
        X^{kn}
      \end{aligned}
    \end{equation}
    where $0 \leq i <m$ and
    $B:=(b_{jk})$ 
    denotes an $m\times m^{t}$ matrix of polynomials of degree less
    than $n$.  The product $A\cdot B$ can be computed according to
    \short\ref{Matrix:a} in the claimed running time.  Multiplication
    by $X^{kn}$ amounts to mere coefficient shifts rather than
    arithmetic operations.  And observing that $\deg\big((A\cdot
    B)_{ik}\big)<2n$, only two consecutive terms in the right hand
    side of~(\ref{eqSum}) can overlap.  So evaluating this sum amounts
    to $m^{t}$-fold addition of pairs of polynomials of degree less
    than $n$.  Since $\omega(1,1,t) \geq 1+t$ by virtue of
    (\ref{eqOmega}), this last cost of $n m^{1+t}$ is also covered by
    the claimed complexity bound.
  \item[\ref{Comp:a}] Write each $p_{i}$ as a polynomial in $Y$ with
    coefficients from $\bbbK[X]$, \ie{}
    $$
    p_{i}(X,Y) = \sum_{0 \leq j < m} q_{ij}(X)\cdot Y^{j}
    $$
    with all $q_{ij}(X)$ of degree less than $n$.  Iteratively
    compute the $m$ polynomials $g_{j}(X):=g^{j}(X)\rem f(X)$, each of
    degree less than $n m^{t}$, within time~$\softO(n m^{1+t})$ by
    fast division with remainder (see \eg{} Theorem~9.6 in
    \citealt{gatger03}).
    
    By multiplying the matrix $A:=(q_{ij})$ to the vector $b:=(g_{j})$
    according to \ref{Matrix:d}, determine the $m$ polynomials
    $$
    \tilde{p}_{i}(X) := \sum_{0\leq j < m} q_{ij}(X)\cdot g_{j}(X),
    \qquad 0\leq i < m
    $$
    of degree less than $n + n m^{t}$.  For each $i$ reduce again
    modulo $f(X)$ and obtain $p_{i}\big(X,g(X)\big)\rem f(X)$ using
    another $\softO(n m^{1+t})$ operations.
    Since $\omega(1,1,t)\geq 1+t$ according to (\ref{eqOmega}), both
    parts are covered by the claimed running time $\smoothO(n
    m^{\omega(1,1,t)})$.
    \qed%
  \end{enumerate}
\end{proof}

\ref{Matrix} puts us in position to prove \ref{Comp}.

\begin{proof}[\ref{Comp}]
  Without loss of generality we assume that $m$ is a square.  We use a
  baby step, giant step strategy: Partition $p$ into $\sqrt{m}$
  polynomials $p_{i}$ of $\deg_{Y}(p_{i})<\sqrt{m}$, \ie{}
  $$
  p(X,Y)\quad=\quad\sum_{0 \leq i < \sqrt{m}} p_{i}(X,Y)\cdot
  Y^{i\sqrt{m}} \enspace .
  $$
  Then apply \whole\ref{Comp:a} with $t=2$ and $m$ replaced by
  $\sqrt{m}$ to obtain the $\sqrt{m}$ polynomials
  $\tilde{p}_{i}(X):=p_{i}\big(X,g(X)\big)\rem f(X)$ within
  $\smoothO(n m^{\myomega/2})$ operations.  Iteratively determine the
  $\sqrt{m}$ polynomials $\tilde{g}_{i}(X) :=
  \big(g(X)^{\sqrt{m}}\big)^{i}\rem f(X)$ for $0 \leq i <\sqrt{m}$
  within $\softO(n m^{3/2})$.  Again, $\myomega\geq3$ asserts this to
  remain in the claimed bound.  Finally compute
  $$
  p\big(X,g(X)\big)\rem f(X) \quad=\quad \sum_{0 \leq i < \sqrt{m}}
  \Big(\tilde{p}_{i}(X)\cdot\tilde{g}_{i}(X)\Big)\rem f(X)
  $$
  using another time $\softO(n m^{3/2})$.
  \qed%
\end{proof}


Based on \ref{Comp}, the following algorithm realizes the idea
expressed in Section~\ref{secIdea}.
\begin{algorithm}{genericmultieval}[Generic multipoint evaluation of a
  bivariate polynomial]
\item Coefficients of a polynomial $p\in\bbbK[X,Y]$ of
  $\deg_{X}(p)<n$, $\deg_{Y}(p)<m$ and points $(x_{k},y_{k})$ for
  $0\leq k < n m$ with pairwise different first coordinates $x_{k}$.
\item The values $p(x_{k},y_{k})$ for $0 \leq k <n m$.
\item\algolabel{f} Compute the univariate polynomial $\displaystyle
  f(X):=\prod_{0 \leq k< n m} (X-x_{k})\in\bbbK[X]$.
\item\algolabel{g} Compute an interpolation polynomial $g\in\bbbK[X]$
  of degree less than $n m$ satisfying $g(x_{k})=y_{k}$ for all $0
  \leq k <n m$.
\item\algolabel{comp} Apply \ref{Comp} to obtain $\tilde{p}(X) :=
  p\big(X,g(X)\big)\rem f(X)$.
\item\algolabel{uni} Multi-evaluate this univariate polynomial
  $\tilde{p}\in\bbbK[X]$ of degree less than $n m$ at the $n m$
  arguments $x_{k}$.
\item \RETURN $(\tilde{p}(x_{k}))_{0\leq k <n m}$.
\end{algorithm}

\begin{proof}[\ref{Main2}]
  The algorithm is correct by construction.
  
  \stepref{genericmultieval}{f} can be done in $\softO(n m)$
  arithmetic operations.
  As the points $(x_{k},y_{k})$ have pairwise different first
  coordinates, the interpolation problem in
  Step~\ref{genericmultieval-g} is solvable and, by virtue of
  \ref{Unibasics:c}, in running time $\softO(n m)$.
  For Step~\ref{genericmultieval-comp} \ref{Comp} guarantees running
  time $\smoothO(n m^{\myomega/2})$.
  According to \ref{Unibasics:b}, Step~\ref{genericmultieval-uni} is
  possible within time $\softO(n m)$.
  Summing up, we obtain the claimed running time.
  \qed%
\end{proof}


\section{Evaluating at degenerate points}
\label{secRestriction}

Here we indicate how certain fields $\bbbK$ permit to remove the
condition on the evaluation point set imposed in \ref{Main2}.  
The idea is to rotate or shear
the situation slightly, so that afterwards the point set has pairwise
different first coordinates.  To this end choose $\theta \in \bbbK$
arbitrary such that
\begin{equation} \label{eqTheta}
  \#\Set{
    x_{k}+\theta y_{k}; 0\leq k < N } \quad =\quad  N
\end{equation}
where $N:=nm$ denotes the number of points.  Then replace each
$(x_k,y_k)$ by $(x'_{k},y'_{k}) := (x_{k}+\theta y_{k}, y_{k})$ and
the polynomial $p$ by $\hat{p}(X,Y) := p(X-\theta Y, Y)$.  This can be
done with $\softO\left(n^{2}+m^{2}\right)$ arithmetic operations, see
the more general \ref{Subst} below.  In any case a perturbation like
this might even be a good idea if there are points whose first
coordinates are `almost equal' for reasons of numerical stability.
\begin{lemma}
  \label{Generic}
  Let $\bbbK$ denote a field and $P = \Set{ (x_k,y_k)\in\bbbK^2; 0\leq
    k <N }$ a collection of $N$ planar points.
  \begin{enumerate}
  \item
    \label{Generic:a}
    If $\#\bbbK\geq N^2$, then $\theta\in\bbbK$ chosen uniformly at
    random satisfies (\ref{eqTheta}) with probability at least
    $\tfrac{1}{2}$.
    Using $\bigO(\log N)$ guesses and a total of $\bigO(N\cdot\log^2 N)$ 
    operations, we can thus find an appropriate $\theta$
    with high probability.

    If $\bbbK$ is even infinite, a single guess almost certainly
    suffices.
  \item
    \label{Generic:b}
    In case $\bbbK=\bbbR$ or $\bbbK=\bbbC$, we can deterministically
    find an appropriate $\theta$ in time $\bigO(N\cdot\log N)$.
  \item
    \label{Generic:c}
    For a fixed proper extension field $\bbbL$ of $\bbbK$,
    any $\theta \in \bbbL \setminus \bbbK$ will do.
  \end{enumerate}
\end{lemma}
Applying \ref{Generic:a} or \ref{Generic:b} together with \ref{Subst}
affects the running time of \ref{Main2} only by the possible change in
the $Y$-degree.  Using \ref{Generic:c} means that all subsequent
computations must be performed in $\bbbL$.  This increases all further
costs by no more than an additional constant factor depending on the
degree $[\bbbL:\bbbK]$ only.

\begin{proof}
  \begin{itemize}
  \item[\short\ref{Generic:a}] Observe that an undesirable $\theta$
    with $x_{k}+\theta y_{k}=x_{k'}+\theta y_{k'}$ implies
    $y_k=y_{k'}$ or $\theta=\frac{x_k-x_{k'}}{y_{k'}-y_k}$.  In the
    latter case, $\theta$ is thus uniquely determined by $\{k,k'\}$.
    Since there are at most $\binom{N}{2}<N^2/2$ such choices
    $\{k,k'\}$, no more than half of the $\#\bbbK\geq N^{2}$ possible
    values of $\theta$ can be undesirable.
  \item[\short\ref{Generic:b}] If $\bbbK = \bbbR$ choose
    $\theta>0$ such that $\theta \cdot
    (y_{\textnormal{max}}-y_{\textnormal{min}}) <
    \min\Set{x_{k}-x_{k'};x_k>x_{k'}}$.  Such a value~$\theta$ can be
    found in linear time after sorting the points with respect to
    their $x$-coordinate.

    In case $\bbbK = \bbbC$, we can do the same with
    respect to the real parts.
  \item[\short\ref{Generic:c}] Simply observe that $1$ and $\theta$
    are linearly independent.
    \qed%
  \end{itemize}
\end{proof}

We now state the already announced
\begin{lemma}
  \label{Subst}
  Let $R$ be a commutative ring with one.
  Given $n\in\bbbN$ and the $n^{2}$ coefficients of a polynomial
  $p(X,Y) \in R[X,Y]$ of degree less than $n$ in both $X$ and $Y$.
  Given furthermore a matrix $A \in R^{2\times 2}$ and a vector $b\in
  R^{2}$.  From this, we can compute the coefficients of the affinely
  transformed polynomial
  $p(a_{11}X+a_{12}Y+b_{1},a_{21}X+a_{22}Y+b_{2})$ using $\bigO(n^{2}
  \cdot \log^{2} n \cdot \loglog n)$ or $\softO(n^{2})$ arithmetic
  operations over $R$.
  
  In the special case $R=\bbbC$ we can decrease the running time to
  $\bigO(n^2 \log n)$.
\end{lemma}
\ref{Subst} straight-forwardly generalizes to $d$-variate polynomials
and $d$\=dimensional affine transformations being applicable within
time $\softO(n^{d})$ for fixed $d$.

\begin{proof}
  We prove this in several steps.
  \begin{itemize}
  \item
    \label{Shift:a}
    First we note that, over any commutative ring $S$ with one, we
    can compute the \emph{\person{Taylor} shift} $p(X+a)$ of a
    polynomial $p \in S[X]$ of degree less than $n$ by an element
    $a\in S$ using $\bigO(n \cdot \log^{2} n \cdot \loglog n)$
    arithmetic operations in $S$.
    
    There are many solutions for computing the \emph{\person{Taylor}
      shift} of a polynomial.  We would like to sketch the divide and
    conquer solution from Fact~2.1(iv) in \cite{gat90c} that works
    over any ring $S$: Precompute all powers $(X+a)^{2^{i}}$ for $0
    \leq i \leq \nu := \floor{ \log_{2} n }$. Then recursively split
    $p(X) = p_{0}(X) + X^{2^{\nu}} p_{1}(X)$ with $\deg p_{0} <
    2^{\nu}$ and calculate $p(X+a) = p_{0}(X+a) + (X+a)^{2^{\nu}}
    p_{1}(X+a)$.  This amounts to $\bigO(n \cdot \log^{2} n \cdot
    \loglog n)$ multiplications in $S$ and $\bigO(n \log n)$ other
    operations.  So we achieve this over any ring $S$ with $\bigO(n
    \cdot \log^{2} n \cdot \loglog n)$ operations.
  \item 
    \label{Shift:b}\label{Shift:c}
    Next let $S=R[Y]$.  Then we can use the previous to compute
    $p(X+a,Y)$ or $p(X+aY,Y)$ for a polynomial $p \in R[X,Y] = S[X]$
    of maximum degree less than $n$ and an element $a \in R$.  Using
    \person{Kronecker} substitution for the multiplications in
    $R[X,Y]$ this can be done with $\bigO(n^2 \cdot \log^{2} n \cdot
    \loglog n)$ arithmetic operations in $R$.
  \item Now we prove the assertion.  Scaling is easy: $p(x,y)\mapsto
    p(\alpha x,y)$ obviously works within $\bigO(n^{2})$ steps.  Use
    this and the discussed shifts once or twice.
  \end{itemize}
  The solution to Problem~2.6 in \cite{binpan94} allows to save a
  factor $\log n \cdot \loglog n$ when $R=S=\bbbC$.
  \qed%
\end{proof}

\section{Conclusion and Further Questions}
\label{secConclusion}

We lowered the upper complexity bound for multi-evaluating dense
bivariate polynomials of degree less than $n$ with $n^{2}$
coefficients at $n^{2}$ points with pairwise different first
coordinates from na{\"{\i}}ve $\bigO(n^{4})$ and $\softO(n^{3})$ to
$\bigO(n^{2.667})$.  The algorithm is based on fast univariate
polynomial arithmetic together with fast matrix multiplication and
will immediately benefit from any future improvement of the latter.

With the same technique, evaluation of a trivariate polynomial of
maximum degree less than $n$ at $n^3$ points can be accelerated from
na{\"{\i}}ve $\bigO(n^{6})$ to $\bigO(n^{4.334})$.

Regarding that the matrix multiplication method of \cite{huapan98} has
huge constants hidden in the big-Oh notation, it might in practice be
preferable to use either the na{\"{\i}}ve $2m^3$ or Strassen's $4.7
m^{2.81}$ algorithm (with some tricks).  Applying them to our approach
still yields bivariate multipoint evaluation within time $\bigO(n^{3})$ or
$\bigO(n^{2.91})$, respectively, with small big-Oh constants and no
hidden factors $\log n$ in the leading term, that is, faster than
\ref{Gridextension}.

Further questions to consider are:
\begin{itemize}
\item Is it possible to remove even the divisions?  This would give a
  much more stable algorithm and it would also
  work over many rings.
\item As $\omega\geq2$, the above techniques will never get below
  running times of order $n^{2.5}$.  Can we achieve an upper
  complexity bound as close as $\softO(n^{2})$ to the information
  theoretic lower bound?
\item Can multipoint evaluation of trivariate polynomials
  $p(X_{1},X_{2},X_{3})$ be performed in time $\smallo(n^{4})$?
\end{itemize}

\paragraph{Acknowledgements}
The authors wish to thank David \cite{epp04} for an inspiring
suggestion that finally led to \ref{Generic:b}.

\newcommand{\hide}[1]{#1}%
\newcommand{\Hide}[1]{#1}%
\bibliography{nusken}

\end{document}